\documentclass[12pt]{article}
\usepackage{ctex}
\usepackage{amssymb,amsmath,color,amsmath,bm}
\usepackage[colorlinks,linkcolor=blue]{hyperref}
\usepackage{geometry}
\usepackage{tabularx}
\usepackage{bm}
\usepackage{authblk}
\usepackage[square, comma, sort&compress, numbers]{natbib}

		\title{\bf Self-orthogonal generalized twisted Reed-Solomon codes}	
		\author{\small Canze Zhu}
	\author{\small Qunying Liao
		\thanks{Corresponding author.\\
			{E-mail. qunyingliao@sicnu.edu.cn (Q. Liao), ~canzezhu@163.com (C. Zhu).}	\\				
			{~Supported by National Natural Science Foundation of China (Grant No. 12071321).}}
		}
	\affil[]{\small (College of Mathematical Science, Sichuan Normal University, Chengdu Sichuan, 610066)}
	\date{}
	\newtheorem{theorem}{Theorem}[section]
	\newtheorem{definition}{Definition}[section]
	\newtheorem{lemma}{Lemma}[section]

	\newtheorem{corollary}{Corollary}[section]
	\newtheorem{remark}{Remark}[section]

	\catcode`@=11 \@addtoreset{equation}{section} \catcode`@=12
\begin{document}
	\maketitle
	{\bf Abstract.}
 {\small In this paper,  by calculating the  dual code of  the Schur square  for the standard twisted  Reed-Solomon code, we give a sufficient and necessary condition for the generalized twisted  Reed-Solomon code with $h+t\le k-1$ to be self-orthogonal, where $k$ is dimension, $h$ is hook and $t$ is twist. And then, we show that there is no self-orthogonal  generalized twisted  Reed-Solomon code under some conditions. Furthermore, several classes of self-orthogonal  generalized twisted  Reed-Solomon codes are constructed, and some of these codes are non-GRS self-orthogonal MDS codes or NMDS codes.
		}\\
	
	{\bf Keywords.}	{\small   generalized twisted Reed-Solomon code;  the Schur product;  non-GRS self-orthogonal MDS code; self-orthogonal NMDS code.}
	
	\section{Introduction}
	An $[n,k,d]$ linear code $\mathcal{C}$ over $\mathbb{F}_q$ is a $k$-dimensional subspace of $\mathbb{F}_q^n$ with minimum (Hamming) distance $d$ and length $n$. For $\mathbf{a}=(a_1,\ldots,a_n)$, $\mathbf{b}=(b_1,\ldots,b_n)$ $\in\mathbb{F}_q^n$, the Euclidean inner product is defined as $\langle \mathbf{a},\mathbf{b}\rangle_E=\sum_{i=1}^{n}a_ib_i$. And then 
	the dual code of $\mathcal{C}$ is defined as $$\mathcal{C}^{\perp}=\{\mathbf{c}^{'}\in\mathbb{F}_q^n~|~\langle\mathbf{c}^{'},\mathbf{c}\rangle_E=0, \text{for any}~\mathbf{c}\in\mathcal{C}\}.$$ We called $\mathcal{C}$ is self-orthogonal if $\mathcal{C}\subseteq\mathcal{C}^{\perp}$. Especially, $\mathcal{C}$ is self-dual if $\mathcal{C}=\mathcal{C}^{\perp}$.
	
	Both self-orthogonal codes and  MDS codes have widely practical applications. On the one hand, in quantum information theory,   
	self-orthogonal codes can be used to construct pure additive quantum codes \cite{SA99}, quantum stabilizer codes \cite{KA2006},
	and so on. On the other hand, the MDS code can correct maximal number of errors for a given code rate \cite{17}. In addition, MDS codes are closely connected to combinatorial designs and finite geometry \cite{17,28}. For the above reasons, self-orthogonal MDS 
	codes have been attracted much attentions \cite{FX20,HZ21}. As a special case, the self-dual MDS code is of particular importance due to
	its optimality and self-duality, a lot of these codes were constructed in various ways \cite{4,11,12,13,14,15,20,21,33,34,37}, especially, generalized Reed-Solomon (in short, GRS) codes are a class of MDS codes and a lot of self-dual MDS codes are constructed from GRS codes \cite{10,19,25,39,40}.
	
	The near MDS (in short, NMDS) code is introduced in \cite{7}. A linear code $\mathcal{C}$ is called an near MDS code if both $\mathcal{C}$ and its dual code have one singleton defect from being an MDS code. NMDS codes approximate maximal minimum Hamming distance for a given code rate. Futhermore, NMDS codes also have been applicated in secret sharing scheme \cite{35}, the index coding problem \cite{42}, the informed source coding problem \cite{43}, and so on. From both theoretical and practical point of views, to study self-orthogonal NMDS is interesting. 
	In 2017,  inspired by twisted Gabidulin codes \cite{t26}, Beelen, et al. firstly introduced the twisted Reed-Solomon (in short, TRS) codes,  and also showed that TRS codes could be well decoded \cite{2,B}.  Different from GRS codes, a  generalized twisted Reed-Solomon  (in short, GTRS) code is not necessarily MDS \cite{2}. The authors also showed that some of GTRS codes are not GRS codes \cite{3,26}. Furthermore, basing on GTRS codes, Lavauzelle, et al. presented an efficient key-recovery attack used in the McEliece cryptosystem \cite{24}. GTRS codes are also used to construct linear complementary dual (in short, LCD) MDS codes by their applications in cryptography \cite{16,26}. In 2021, Liu et, al. gave the parity-check matrix for the GTRS code with dimension $k$, hook $k-1$ and twist $1$, and then constructed several classes of self-dual MDS or NMDS codes from these GTRS codes \cite{t0}. Recently, the authors generalized Liu et, al.'s work,  
	 presented the parity-check matrix for the GTRS code with any given dimension $k$, hook $h$ and twist $t$,  gave a sufficient and necessary condition for the GTRS code with $h+t=k$ and $h\ge t$ to be self-dual, and then also constructed some self-dual codes with small defect \cite{ZC21}.
	
	In this paper,  by calculating the  dual code of  the Schur square  for the standard TRS code, a sufficient and necessary condition for the GTRS code with $h+t\le k-1$ to be self-orthogonal is given. And then, basing on this sufficient and necessary condition, we prove that there is no self-orthogonal  GTRS code under some conditions. Furthermore, we construct several classes of self-orthogonal  GTRS codes, especially, some of these codes are proved to be non-GRS self-orthogonal MDS codes or NMDS codes.
	
	The rest of this paper is organized as follows. In section 2, some basic notations and results about GTRS codes, GRS codes and the Shur product are given. In section 3,  the  dual code of the Schur square for the standard TRS code is given. In section 4, a sufficient and necessary condition for the GTRS code with $h+t\le k-1$ to be self-orthogonal is obtained, and then several classes of self-orthogonal  GTRS codes are constructed. In section 5, we conclude the whole paper and give the further study.
	
	\section{Preliminaries}
	Throughout this paper, we fix some nonations as follows for convinience.\\
	$\bullet$ $q$ is a power of prime.\\	
	$\bullet$  $\mathbb{F}_q$ is the finite field with $q$ elements, and $\mathbb{F}_q^{*}=\mathbb{F}_q\backslash\{0\}$.\\
	$\bullet$  $\mathbb{F}_q[x]$ is the polynominal ring over $\mathbb{F}_q$.\\
 	$\bullet$  $\alpha_1,\ldots,\alpha_q$ are all the elements in $\mathbb{F}_q$, namely, $\mathbb{F}_q=\{\alpha_1,\ldots,\alpha_q\}$.\\
	$\bullet$  $t$, $k$ and $n$ are positive integers with $3\le k<n\le q$, $h$ is a non-negative integer.\\
	$\bullet$ $i_1,\ldots,i_n$ are positive integers with $1\le i_1<\ldots<i_n\le q$.\\
	$\bullet$ For any positive integer $l$, $\mathbf{\mathcal{S}}_{l}=\{0,1,\ldots,l\}.$\\
	$\bullet$ For any nonempty set $M\subsetneq\{0,1,\ldots,q-1\}$, denote 
	\begin{align*}
	M^{\perp}=\{0,1,\ldots,q-1\}\backslash\{q-1-a\big|a\in M\}.
	\end{align*}
	$\bullet$ $\mathbf{0}=(0,\ldots,0)$, where $0$ is the zero element  in $\mathbb{F}_q$.\\	
	$\bullet$ $\mathbf{1}=(1,\ldots,1)$, where $1$ is the  identity in $\mathbb{F}_q$.\\
	$\bullet$ For any positive integer $s$ and $\boldsymbol{\alpha}=(\alpha_{i_1},\ldots,\alpha_{i_n})$, denote $\boldsymbol{\alpha}^{0}=\mathbf{1}~ \text{and}~\boldsymbol{\alpha}^{s}=(\alpha_{i_1}^s,\ldots,\alpha_{i_n}^s).$\\
	$\bullet$ For any $\boldsymbol{v}_1,\ldots,\boldsymbol{v}_l\in \mathbb{F}_{q}^n$, $\langle \boldsymbol{v}_1,\ldots,\boldsymbol{v}_l\rangle$  is the $\mathbb{F}_{q}$-linear space generated by the set $\{\boldsymbol{v}_1,\ldots,\boldsymbol{v}_l\}$.\\
	\subsection{GTRS codes}
	In this subsection, we give some notations and lemmas about GTRS codes and GRS codes, respectively.
	\begin{definition}
		For   a $\mathbb{F}_{q}$-linear subspace $\mathcal{V}$ of $\mathbb{F}_q[x]$, let $\boldsymbol{\alpha}=(\alpha_{i_1},\ldots,\alpha_{i_n})$ and $\boldsymbol{v}=(v_1,\ldots,v_n)\in(\mathbb{F}_q^{*})^{n}$, then the linear code with length $n$ is defined as
		\begin{align*}
		\mathcal{C}_{\mathcal{V}}(\boldsymbol{\alpha},\boldsymbol{v})=\{(v_1f({\alpha_1}),\ldots,v_nf(\alpha_n))~|~ f(x)\in \mathcal{V} \}.
		\end{align*}
	\end{definition}

	\begin{remark}\label{r1}
	  (1) For $i=1,\ldots,k$ and  $f_i(x)\in\mathbb{F}_q[x]$, let $$\mathcal{V}=\big\{a_1f_1(x)+\cdots+a_kf_k(x)~|~a_i\in\mathbb{F}_q~(i=1,\ldots,k)\big\}.$$ By the definition of $\mathcal{C}_{\mathcal{V}}(\boldsymbol{\alpha},\boldsymbol{v})$, we have
	\begin{align*}
	\mathcal{C}_{\mathcal{V}}(\boldsymbol{\alpha},\boldsymbol{v})=\big\langle 	\big(v_1f_i({\alpha_1}),\ldots,v_nf_i(\alpha_n)\big)~(i=1,\ldots,k)\big\rangle.
	\end{align*}
	
	(2) For convinience, we denote $\mathcal{C}_\mathcal{V}(\boldsymbol{v})=\mathcal{C}_{\mathcal{V}}(\boldsymbol{\alpha},\boldsymbol{v})$ and $\mathcal{C}_\mathcal{V}=\mathcal{C}_{\mathcal{V}}(\boldsymbol{\alpha},\boldsymbol{1})$ for $\boldsymbol{\alpha}=(\alpha_1,\ldots,\alpha_q)$.
	\end{remark}
	\begin{definition}\cite{2}
	Let $h\le k-1$ and $\eta\in\mathbb{F}_q^{*}$. Define the set of $(k,t,h,\eta)$-twisted polynomials as 
		\begin{align*}
			\mathcal{V}_{k,t,h,\eta}=\bigg\{f(x)=\sum_{i=0}^{k-1}a_ix^i+\eta a_hx^{k-1+t}~|~a_i\in\mathbb{F}_q~(i=0,\ldots,k-1)\bigg\},
		\end{align*}
		which is a $k$-dimensional $\mathbb{F}_q$-linear subspace of $\mathbb{F}_q[x]$. We call $h$ hook and $t$ twist.
	\end{definition}
	
	\begin{definition}\cite{2}
 For  $h\le k-1$ and $k+t\le n$, let $\eta\in\mathbb{F}_q^{*}$, $\boldsymbol{\alpha}=(\alpha_{i_1},\ldots,\alpha_{i_n})$ and $\boldsymbol{v}=(v_1,\ldots,v_n)\in(\mathbb{F}_q^{*})^{n}$, then $\mathcal{C}_{\mathcal{V}_{k,t,h,\eta}}(\boldsymbol{\alpha},\boldsymbol{v})$ is called the GTRS code. Especially,  $\mathcal{C}_{\mathcal{V}_{k,t,h,\eta}}(\boldsymbol{\alpha},\boldsymbol{1})$ is called the TRS code.
	\end{definition}
\begin{remark}
	For $\boldsymbol{\alpha}=(\alpha_{1},\ldots,\alpha_{q})$, $\mathcal{C}_{\mathcal{V}_{k,t,h,\eta}}=\mathcal{C}_{\mathcal{V}_{k,t,h,\eta}}(\boldsymbol{\alpha},\boldsymbol{1})$ is the standard TRS code. 
\end{remark}
	\begin{remark}\label{r5}
		By the definition of $\mathcal{C}_{\mathcal{V}_{k,t,h,\eta}}(\boldsymbol{\alpha},\boldsymbol{1})$, it is easy to see that
		\begin{align*}
		\mathcal{C}_{\mathcal{V}_{k,t,h,\eta}}(\boldsymbol{\alpha},\boldsymbol{1})=\big\langle\boldsymbol{\alpha}^{s},\boldsymbol{\alpha}^{h}+\eta\boldsymbol{\alpha}^{k-1+t}(s\in\mathbf{\mathcal{S}}_{k-1}\backslash\{h\})\big\rangle.
		\end{align*}
	\end{remark}
	\begin{remark}
		For  any  $f(x)\in \mathcal{V}_{k,t,h,\eta}$, we have  $\deg f(x)\le k-1+t<n$, thus  $\mathcal{C}_{\mathcal{V}_{k,t,h,\eta}}(\boldsymbol{\alpha},\boldsymbol{v})$ is with dimension $k$ and minimum distance $d\ge n-k-t+1$.
	\end{remark}
	
	\begin{lemma}[Lemma 1, \cite{16}]\label{mnds} For $\boldsymbol{\alpha}=(\alpha_{i_1},\ldots,\alpha_{i_n})$ and $\eta\in\mathbb{F}_q^{*}$, let $$T_k(\boldsymbol{\alpha})=\Big\{(-1)^{k}\prod_{i\in I}\alpha_i^{-1}~\bigg|~\forall I\subsetneq \{1,\ldots,n\} \text{~with~}|I|=k\Big\}.$$ Then we have
		
		$(1)$	$\mathcal{C}_{\mathcal{V}_{k,1,0,\eta}}(\boldsymbol{\alpha},\boldsymbol{v})$ is an MDS code if and only if $\eta\notin T_k(\boldsymbol{\alpha})$,
		
		$(2)$	$\mathcal{C}_{\mathcal{V}_{k,1,0,\eta}}(\boldsymbol{\alpha},\boldsymbol{v})$ is an NMDS code if and only if $\eta\in T_k(\boldsymbol{\alpha})$.
	\end{lemma}
	
	\begin{lemma}	[Lemma 12, \cite{B}]\label{nmds}
	Let $q$ be odd and $3 \le k \le \frac{q-1}{2}-2$. If $n>\frac{q+1}{2}$, then $\mathcal{C}_{\mathcal{V}_{k,1,0,\eta}}(\boldsymbol{\alpha},\boldsymbol{v})$ is not MDS.
	\end{lemma}
    
    \begin{lemma}[Theorem 17, \cite{2}] \label{lmds}
	Let $\mathbb{F}_s$ be a proper subfield of $\mathbb{F}_q$ and $\texorpdfstring{\bm{\alpha}} ~\in\mathbb{F}_s^{n}$. If $\eta\in \mathbb{F}_q\backslash \mathbb{F}_s$, then  	$\mathcal{C}_{\mathcal{V}_{k,t,h,\eta}}(\boldsymbol{\alpha},\boldsymbol{v})$ is MDS.
    \end{lemma}
\begin{definition}\cite{28}
	Let $\mathcal{V}_k=\{f(x)\in\mathbb{F}_q[x]|\deg f(x)\le k-1\}$, then $\mathcal{C}_{\mathcal{V}_{k}}(\boldsymbol{\alpha},\boldsymbol{v})$ is called the GRS  code. Especially,  $\mathcal{C}_{\mathcal{V}_{k}}(\boldsymbol{\alpha},\boldsymbol{1})$ is called the RS code.
\end{definition}
\begin{remark}
\label{rs}
(1)	By the definition of $\mathcal{C}_{\mathcal{V}_{k,t,h,\eta}}(\boldsymbol{\alpha},\boldsymbol{1})$, it is easy to see that
	\begin{align*}
	\mathcal{C}_{\mathcal{V}_{k}}(\boldsymbol{\alpha},\boldsymbol{1})=\big\langle\boldsymbol{\alpha}^{s}~(s\in\mathbf{\mathcal{S}}_{k-1}\backslash\{h\})\big\rangle.
	\end{align*}
	
(2)	For convinience, we denote $\mathcal{RS}_k=\mathcal{C}_{\mathcal{V}_{k}}(\boldsymbol{\alpha},\boldsymbol{1})$ for $\boldsymbol{\alpha}=(\alpha_1,\ldots,\alpha_q)$.
\end{remark}
\begin{lemma}[Lemma 2.3, \cite{19}]\label{RS}
	$\mathcal{RS}_k^{\perp}=\mathcal{RS}_{q-k}$.
\end{lemma}
	\subsection{The Schur product}
	 The Schur product  is defined as follows.
	\begin{definition}
		Let $\mathbf{x}=(x_1,\ldots,x_n ), \mathbf{y}= (y_1 ,\dots, y_n)\in \mathbb{F}_q^{n}$,  the Schur product  of $\mathbf{x}$ and $\mathbf{y}$ is
		defined as $$x*y:=(x_1y_1,\ldots,x_ny_n),$$  The Schur product  of two $q$-ary codes $\mathcal{C}_1$ and $\mathcal{C}_2$ with length $n$ is
		defined as
		\begin{align*}
		\mathcal{C}_1*\mathcal{C}_2=\langle \mathbf{c}_1*\mathbf{c}_2~|~\mathbf{c}_1\in\mathcal{C}_1,\mathbf{c}_2\in\mathcal{C}_2\rangle.
		\end{align*}
		Especially, for a code $\mathcal{C}$, we call $\mathcal{C}^2:=\mathcal{C}*\mathcal{C}$ the Schur square for $\mathcal{C}$.
	\end{definition} 
	
	\begin{remark}For any two linear codes  $\mathcal{C}_1=\langle \boldsymbol{v}_1,\ldots,\boldsymbol{v}_{k_1}\rangle$ and   $\mathcal{C}_2=\langle \boldsymbol{w}_1,\ldots,\boldsymbol{w}_{k_2}\rangle$ with $\boldsymbol{v}_i,\boldsymbol{w}_j\in\mathbb{F}_q^{n}~(i=1,\ldots,k_1,~j=1,\ldots,k_2)$, then by the definition of  the Schur product, we have
		 \begin{align}\label{S1}
		\mathcal{C}_1*\mathcal{C}_2=\langle \boldsymbol{v}_i*\boldsymbol{w}_j~(i=1,\ldots,k_1,~j=1,\ldots,k_2)\rangle.
		\end{align}
	\end{remark}
	\begin{remark}\label{rsp}
		For any $k\le\frac{n+1}{2}$, $\boldsymbol{\alpha}$ and $\boldsymbol{v}$, it is easy to see that  
		$\mathcal{C}_{\mathcal{V}_k}^2(\boldsymbol{\alpha},\boldsymbol{1})=\mathcal{C}_{\mathcal{V}_{2k-1}}(\boldsymbol{\alpha},\boldsymbol{1})$, and then $$\dim\big(\mathcal{C}_{\mathcal{V}_{k}}^2(\boldsymbol{\alpha},\boldsymbol{v})\big)=\dim\big(\mathcal{C}_{\mathcal{V}_{k}}^2(\boldsymbol{\alpha},\boldsymbol{1})\big)=2k-1.$$
	\end{remark}

	\section{ $\mathcal{C}^2_{\mathcal{V}_{k,t,h,\eta}}$ and Its Dual Code}
	
	$\mathcal{C}^2_{\mathcal{V}_{k,t,h,\eta}}(\boldsymbol{\alpha},\boldsymbol{1})$ and $\mathcal{C}^2_{\mathcal{V}_{k,t,h,\eta}}$  are given in Lemmas \ref{L1}-\ref{L2}, respectively.
	\begin{lemma}\label{L1} For $t+h\le k-1$, $\eta\in\mathbb{F}_{q}^{*}$ and $\boldsymbol{\alpha}=(\alpha_{i_1},\ldots,\alpha_{i_n})$, we have{\small
		\begin{align*}
			&\mathcal{C}^2_{\mathcal{V}_{k,t,h,\eta}}(\boldsymbol{\alpha},\boldsymbol{1})\\
			=&\begin{cases}
				\big\langle\boldsymbol{\alpha}^{s},\boldsymbol{\alpha}+\eta\boldsymbol{\alpha}^{2k-1},\boldsymbol{1}+\eta^2\boldsymbol{\alpha}^{4k-4}~(s\in\mathbf{\mathcal{S}}_{3k-3}\backslash\{0,1,2k-1\})\big\rangle,
				& \text{if~}h=0\text{~and~}t=k-1;\\
				\big\langle\boldsymbol{\alpha}^{s},\boldsymbol{1}+\eta^2\boldsymbol{\alpha}^{2k+2t-2}~(s\in\mathbf{\mathcal{S}}_{2k-2+t}\backslash\{0\})\big\rangle,&\text{if~}h=0\text{~and~}t\le k-2; \\
				\big\langle\boldsymbol{\alpha}^{s}~(s\in\mathbf{\mathcal{S}}_{2k-2+t}\cup\{2k+2t-2\})\big\rangle,&\text{if~} h\ge 1\text{~and~}t+h\le k-1.\\
			\end{cases}
		\end{align*} }
	\end{lemma}

	{\bf Proof}. By Remarks \ref{r1} and \ref{rsp}, one has
	{
	\begin{align*}
	\begin{aligned}
	&\mathcal{C}^2_{\mathcal{V}_{k,t,h,\eta}}(\boldsymbol{\alpha},\boldsymbol{1})\\
	=&\big\langle\boldsymbol{\alpha}^i*\boldsymbol{\alpha}^j,\boldsymbol{\alpha}^l*(\boldsymbol{\alpha}^h+\eta\boldsymbol{\alpha}^{k-1+t}),(\boldsymbol{\alpha}^h+\eta\boldsymbol{\alpha}^{k-1+t})*(\boldsymbol{\alpha}^h+\eta\boldsymbol{\alpha}^{k-1+t})(i,j,l\in\mathbf{\mathcal{S}}_{k-1}\backslash\{h\})\big\rangle\\
	=&\big\langle\boldsymbol{\alpha}^{i+j},\boldsymbol{\alpha}^{h+l}+\eta\boldsymbol{\alpha}^{k-1+t+l},\boldsymbol{\alpha}^{2h}+2\eta\boldsymbol{\alpha}^{k-1+t+h}+\eta^2\boldsymbol{\alpha}^{2k+2t-2}~(i,j,l\in\mathbf{\mathcal{S}}_{k-1}\backslash\{h\})\big\rangle.
	\end{aligned}
	\end{align*}}
	
	By the assumption $t+h\le k-1$, we have the following five cases.\\
	
	{\bf Case 1}. If $h=0$ and $t= k-1$, then 
	{\small	\begin{align*}
		&\mathcal{C}^2_{\mathcal{V}_{k,t,h,\eta}}(\boldsymbol{\alpha},\boldsymbol{1})\\=&\big\langle\boldsymbol{\alpha}^{s},\boldsymbol{\alpha}^{l}+\eta\boldsymbol{\alpha}^{2k-2+l},\boldsymbol{1}+2\eta\boldsymbol{\alpha}^{2k-2}+\eta^2\boldsymbol{\alpha}^{4k-4}~(s\in\{2,\ldots,2k-2\},~l\in\{1,\ldots,k-1\})\big\rangle\\
		=&\big\langle\boldsymbol{\alpha}^{s},\boldsymbol{\alpha}+\eta\boldsymbol{\alpha}^{2k-1},\boldsymbol{\alpha}^{2k-2+l},\boldsymbol{1}+\eta^2\boldsymbol{\alpha}^{4k-4}~(s\in\{2,\ldots,2k-2\},~l\in\{2,\ldots,k-1\})\big\rangle\\
		=&\big\langle\boldsymbol{\alpha}^{s},\boldsymbol{\alpha}^{l},\boldsymbol{\alpha}+\eta\boldsymbol{\alpha}^{2k-1},\boldsymbol{1}+\eta^2\boldsymbol{\alpha}^{4k-4}~(s\in\{2,\ldots,2k-2\},~l\in\{2k,\ldots,3k-3\})\big\rangle\\
		=&\big\langle\boldsymbol{\alpha}^{s},\boldsymbol{\alpha}+\eta\boldsymbol{\alpha}^{2k-1},\boldsymbol{1}+\eta^2\boldsymbol{\alpha}^{4k-4}~(s\in\mathbf{\mathcal{S}}_{3k-3}\backslash\{0,1,2k-1\})\big\rangle.
		\end{align*}}
	
	{\bf Case 2}. If $h=0$ and $t\le k-2$, then 
{\small	\begin{align*}
		&\mathcal{C}^2_{\mathcal{V}_{k,t,h,\eta}}(\boldsymbol{\alpha},\boldsymbol{1})\\=&\big\langle\boldsymbol{\alpha}^{s},\boldsymbol{\alpha}^{l}+\eta\boldsymbol{\alpha}^{k+t-1+l},\boldsymbol{1}+2\eta\boldsymbol{\alpha}^{k+t-1}+\eta^2\boldsymbol{\alpha}^{2k+2t-2}~(s\in\{2,\ldots,2k-2\},~l\in\{1,\ldots,k-1\})\big\rangle\\
		=&\big\langle\boldsymbol{\alpha}^{s},\boldsymbol{\alpha}+\eta\boldsymbol{\alpha}^{k+t},\boldsymbol{\alpha}^{k+t-1+l},\boldsymbol{1}+\eta^2\boldsymbol{\alpha}^{2k+2t-2}~(s\in\{2,\ldots,2k-2\},~l\in\{2,\ldots,k-1\})\big\rangle\\
		=&\big\langle\boldsymbol{\alpha}^{s},\boldsymbol{\alpha}^{l},\boldsymbol{1}+\eta^2\boldsymbol{\alpha}^{2k+2t-2}~(s\in\{1,2,\ldots,2k-2\},~l\in\{k+1+t,\ldots,2k-2+t\})\big\rangle\\
		=&\big\langle\boldsymbol{\alpha}^{s},\boldsymbol{1}+\eta^2\boldsymbol{\alpha}^{2k+2t-2}~(s\in\mathbf{\mathcal{S}}_{2k-2+t}\backslash\{0\})\big\rangle.
	\end{align*}}

{\bf Case 3}. If $h=1$ and $t\le k-2$, then 
{\small	\begin{align*}
	&\mathcal{C}^2_{\mathcal{V}_{k,t,h,\eta}}(\boldsymbol{\alpha},\boldsymbol{1})\\	=&\big\langle\boldsymbol{\alpha}^{s},\boldsymbol{\alpha}^{l+1}+\eta\boldsymbol{\alpha}^{k+t-1+l},~2\boldsymbol{\alpha}^{k+t}+\eta\boldsymbol{\alpha}^{2k+2t-2}(s\in\{0,2,\ldots,2k-2\},l\in\{0,2,\ldots,k-1\})\big\rangle\\
	=&\big\langle\boldsymbol{\alpha}^{s},\boldsymbol{\alpha}+\eta\boldsymbol{\alpha}^{k+t-1}, \boldsymbol{\alpha}^{k+t-1+l},2\boldsymbol{\alpha}^{k+t}+\eta\boldsymbol{\alpha}^{2k+2t-2}~(s\in\{0,2,\ldots,2k-2\},~l\in\{2,\ldots,k-1\})\big\rangle\\
	=&\big\langle\boldsymbol{\alpha}^{s},\boldsymbol{\alpha}^{l},\boldsymbol{\alpha}^{2k+2t-2}~(s\in\mathbf{\mathcal{S}}_{2k-2},~l\in\{k+1+t,\ldots,2k-2+t\})\big\rangle\\
	=&\big\langle\boldsymbol{\alpha}^{s}~(s\in\mathbf{\mathcal{S}}_{2k-2+t}\cup\{2k+2t-2\})\big\rangle.
	\end{align*}}

{\bf Case 4}. If $2\le h\le k-3$ and $t+h\le k-1$, then 
{\small	\begin{align*}
	&\mathcal{C}^2_{\mathcal{V}_{k,t,h,\eta}}(\boldsymbol{\alpha},\boldsymbol{1})\\	
	=&
	\big\langle\boldsymbol{\alpha}^{s},\boldsymbol{\alpha}^{h+l}+\eta\boldsymbol{\alpha}^{k+t-1+l},~2\boldsymbol{\alpha}^{k+t+h-1}+\eta\boldsymbol{\alpha}^{2k+2t-2}~(s\in\mathbf{\mathcal{S}}_{2k-2},~l\in\mathbf{\mathcal{S}}_{k-1}\backslash\{h\})\big\rangle\\
	=&
	\big\langle\boldsymbol{\alpha}^{s},\boldsymbol{\alpha}^{k+t-1+l},\boldsymbol{\alpha}^{2k+2t-2}~(s\in\mathbf{\mathcal{S}}_{2k-2},~l\in\mathbf{\mathcal{S}}_{k-1}\backslash\{h\})\big\rangle\\
	=&
	\big\langle\boldsymbol{\alpha}^{s},\boldsymbol{\alpha}^{l},\boldsymbol{\alpha}^{2k+2t-2}~(s\in\mathbf{\mathcal{S}}_{2k-2},~l\in\{k+t-1,\ldots,2k-2+t\}\backslash\{k+t+h-1\})\big\rangle\\
	=&\big\langle\boldsymbol{\alpha}^{s}~(s\in\mathbf{\mathcal{S}}_{2k-2+t}\cup\{2k+2t-2\})\big\rangle.
	\end{align*}}

{\bf Case 5}. If $h=k-2$ and $t=1$, then 
	{\small	\begin{align*}
	&\mathcal{C}^2_{\mathcal{V}_{k,t,h,\eta}}(\boldsymbol{\alpha},\boldsymbol{1})\\	=&\big\langle\boldsymbol{\alpha}^{s},\boldsymbol{\alpha}^{k-2+l}+\eta\boldsymbol{\alpha}^{k+l},~2\boldsymbol{\alpha}^{2k-2}+\eta\boldsymbol{\alpha}^{2k}~(s\in\{0,\ldots,2k-4,2k-2\},~l\in\{0,\ldots,k-3,k-1\})\big\rangle\\
	=&
	\big\langle\boldsymbol{\alpha}^{s},\boldsymbol{\alpha}^{k+l},\boldsymbol{\alpha}^{2k-3}+\eta\boldsymbol{\alpha}^{2k-1},\boldsymbol{\alpha}^{2k}~(s\in\mathbf{\mathcal{S}}_{2k-2},~l\in\{0,\ldots,k-3\})\big\rangle\\
	=&\big\langle\boldsymbol{\alpha}^{s}~(s\in\mathbf{\mathcal{S}}_{2k})\big\rangle.
	\end{align*}}
\indent So far, by {\bf Cases} $1$-$5$, we complete the proof. $\hfill\Box$\\

By Remark $\ref{rsp}$, we know that $\mathcal{C}_{\mathcal{V}_{k}}^2(\boldsymbol{\alpha},\boldsymbol{v})=2k-1$ if $k\le\frac{n+1}{2}$, thus by Lemma \ref{L1}, we have the following
\begin{remark}\label{r4}
	 (1) For $(h,t)=(0,k-1)$ with $k<\frac{n-1}{2}$,  $\boldsymbol{\alpha}^2,\ldots, \boldsymbol{\alpha}^{2k-2},  \boldsymbol{\alpha}+\eta\boldsymbol{\alpha}^{2k-1},\boldsymbol{\alpha}^{2k} ,\boldsymbol{\alpha}^{2k+1}$ are $\mathbb{F}_q$-linear independent,  thus $\dim\!\big(\mathcal{C}^2_{\mathcal{V}_{k,t,h,\eta}}(\boldsymbol{\alpha},\boldsymbol{v})\big)\!=\!\dim\!\big(\mathcal{C}^2_{\mathcal{V}_{k,t,h,\eta}}(\boldsymbol{\alpha},\boldsymbol{1})\big)\!\ge 2k,$  and then $\mathcal{C}^2_{\mathcal{V}_{k,t,h,\eta}}(\boldsymbol{\alpha},\boldsymbol{v})$ is non-GRS.
	 
	 (2) For $h=0\text{~and~}$ $2\le t\le k-2$ with  $k<\frac{n}{2}$, $\boldsymbol{\alpha},\boldsymbol{\alpha}^2,\ldots, \boldsymbol{\alpha}^{2k}$ are $\mathbb{F}_q$-linear independent,  thus $\dim\!\big(\mathcal{C}^2_{\mathcal{V}_{k,t,h,\eta}}(\boldsymbol{\alpha},\boldsymbol{v})\big)=\dim\!\big(\mathcal{C}^2_{\mathcal{V}_{k,t,h,\eta}}(\boldsymbol{\alpha},\boldsymbol{1})\big)\ge 2k$,  and then $\mathcal{C}^2_{\mathcal{V}_{k,t,h,\eta}}(\boldsymbol{\alpha},\boldsymbol{v})$ is non-GRS.

	 (3) For $h\ge 1\text{~and~}$ $ t+h\le k-1$ with $k<\frac{n+1}{2}$, $\boldsymbol{1}, \boldsymbol{\alpha}, \ldots, \boldsymbol{\alpha}^{2k-1}$ are $\mathbb{F}_q$-linear independent,  thus $\dim\big(\mathcal{C}^2_{\mathcal{V}_{k,t,h,\eta}}(\boldsymbol{\alpha},\boldsymbol{v})\big)=\dim\big(\mathcal{C}^2_{\mathcal{V}_{k,t,h,\eta}}(\boldsymbol{\alpha},\boldsymbol{1})\big)\ge 2k$,  and then $\mathcal{C}^2_{\mathcal{V}_{k,t,h,\eta}}(\boldsymbol{\alpha},\boldsymbol{v})$ is non-GRS.
\end{remark}

\begin{lemma}\label{L2} For $3\le k\le \frac{q}{2}$, $t+h\le k-1$, and $\eta\in\mathbb{F}_q^{*}$, the following three assertions hold.\\
	
$(1)$ If $\frac{q-t+1}{2}<k\le\frac{q}{2}$, then
	\begin{align*}
	\mathcal{C}^2_{\mathcal{V}_{k,t,h,\eta}}=
	\mathbb{F}_q^{q}.
	\end{align*} 
	
$(2)$ If $\frac{q-2t+1}{2}<k\le \frac{q-t+1}{2}$, then we have
	{\small	\begin{align*}
		&\mathcal{C}^2_{\mathcal{V}_{k,t,h,\eta}}\\
		=&\begin{cases}
		\big\langle\boldsymbol{\alpha}^{s},\boldsymbol{\alpha}+\eta\boldsymbol{\alpha}^{2k-1},\boldsymbol{1}+\eta^2\boldsymbol{\alpha}~(\mathbf{\mathcal{S}}_{3k-3}\backslash\{0,1,2k-1\})\big\rangle,
		&\!\!\!\! \text{if~}q=4k-4,h=0\text{~and~}t=k-1;\\
		\big\langle\boldsymbol{\alpha}^{s},\boldsymbol{\alpha}+\eta\boldsymbol{\alpha}^{2k-1}~(\mathbf{\mathcal{S}}_{3k-3}\backslash\{1,2k-1\})\big\rangle,
		&\!\!\!\! \text{if~}q<4k-4,h=0\text{~and~}t=k-1;\\
		\mathcal{RS}_{2k-1+t},&\!\!\!\!\text{if~}h=0\text{~and~}t\le k-2;\\
		&~\!\text{or~} h\ge 1\text{~and~}t+h\le k-1,
		\end{cases}
		\end{align*} 
	}
and \begin{align}\label{dd1}
	\dim\big(\mathcal{C}^2_{\mathcal{V}_{k,t,h,\eta}}\big)
	=&\begin{cases}
	3k-3,
	&\text{if~}h=0,t=k-1;\\
	2k-1+t,&\text{if~}h=0\text{~and~}t\le k-2,\\&~\text{~or~} h\ge 1\text{~and~}t+h\le k-1.
	\end{cases}
\end{align}
$(3)$ If $3\le k\le \frac{q-2t+1}{2}$, then we have
	{\small
		\begin{align*}
		&\mathcal{C}^2_{\mathcal{V}_{k,t,h,\eta}}\\
		=&\begin{cases}
		\big\langle\boldsymbol{\alpha}^{s},\boldsymbol{\alpha}+\eta\boldsymbol{\alpha}^{2k-1},\boldsymbol{1}+\eta^2\boldsymbol{\alpha}^{4k-4}~(\mathbf{\mathcal{S}}_{3k-3}\backslash\{0,1,2k-1\})\big\rangle,
		& \text{if~}h=0\text{~and~}t=k-1;\\
		\big\langle\boldsymbol{\alpha}^{s},\boldsymbol{1}+\eta^2\boldsymbol{\alpha}^{2k+2t-2}~(s\in\mathbf{\mathcal{S}}_{2k-2+t}\backslash\{0\})\big\rangle,&\text{if~}h=0\text{~and~}t\le k-2; \\
		\big\langle\boldsymbol{\alpha}^{s}~(s\in\mathbf{\mathcal{S}}_{2k-2+t}\cup\{2k+2t-2\})\big\rangle,&\text{if~} h\ge 1\text{~and~}t+h\le k-1,\\
		\end{cases}
		\end{align*} }
	and \begin{align*}
	\dim\big(\mathcal{C}^2_{\mathcal{V}_{k,t,h,\eta}}\big)
	=&\begin{cases}3k-3,
	& \text{if~}h=0\text{~and~}t=k-1;\\
	2k-1+t,&\text{if~}h=0\text{~and~}t\le k-2; \\2k+t,&\text{if~} h\ge 1\text{~and~}t+h\le k-1.\\
	\end{cases}
	\end{align*}
\end{lemma}

{\bf Proof}. For any positive integer $l$, obviously $\boldsymbol{\alpha}^{q-1+l}=\boldsymbol{\alpha}^{l}$,  and then by Lemma \ref{L1}, we have the following three cases.

 {\bf Case 1}. For $\frac{q-t+1}{2}<k\le\frac{q}{2}$, i.e.,
 $$2k-1 \le q-1< 2k-2+t,$$ 
 and then
  $$ t<2k+2t-2-(q-1)\le 2t-1\le 2k-3.$$
Thus
	\begin{align*}
	&\mathcal{C}^2_{\mathcal{V}_{k,t,h,\eta}}\\
	=&\begin{cases}
	\big\langle\boldsymbol{\alpha}^{s},\boldsymbol{\alpha}+\eta\boldsymbol{\alpha}^{2k-1},\boldsymbol{1}+\eta^2\boldsymbol{\alpha}^{4k-4-(q-1)}~(s\in\mathbf{\mathcal{S}}_{q-1}\backslash\{0,2k-1\})\big\rangle,
	& \text{if~}h=0\text{~and~}t=k-1;\\
	\big\langle\boldsymbol{\alpha}^{s},\boldsymbol{1}+\eta^2\boldsymbol{\alpha}^{2k+2t-2-(q-1)}~(s\in\mathbf{\mathcal{S}}_{q-1}\backslash\{0\})\big\rangle,&\text{if~}h=0\text{~and~}t\le k-2; \\
	\big\langle\boldsymbol{\alpha}^{s},\boldsymbol{\alpha}^{2k+2t-2-(q-1)}~(s\in\mathbf{\mathcal{S}}_{q-1})\big\rangle,&\text{if~} h\ge 1\text{~and~}t+h\le k-1.\\
	\end{cases}\\
	=&\mathbb{F}_q^q.
	\end{align*} 
 
 {\bf Case 2}. For $\frac{q-2t+1}{2}<k\le \frac{q-t+1}{2}$, i.e., $$2k-2+t\le q-1<2k+2t-2,$$ and then
  $$ 1\le 2k+2t-2-(q-1)\le t\le 2k-2+t.$$
  Thus
  {\small
  \begin{align*}
  &\mathcal{C}^2_{\mathcal{V}_{k,t,h,\eta}}\\
  =&\begin{cases}
  \big\langle\boldsymbol{\alpha}^{s},\boldsymbol{\alpha}+\eta\boldsymbol{\alpha}^{2k-1},\boldsymbol{1}+\eta^2\boldsymbol{\alpha}^{4k-4-(q-1)}~(\mathbf{\mathcal{S}}_{3k-3}\backslash\{0,1,2k-1\})\big\rangle,
  & \text{if~}h=0\text{~and~}t=k-1;\\
  \big\langle\boldsymbol{\alpha}^{s},\boldsymbol{1}+\eta^2\boldsymbol{\alpha}^{2k+2t-2-(q-1)}~(s\in\mathbf{\mathcal{S}}_{2k-2+t}\backslash\{0\})\big\rangle,&\text{if~}h=0\text{~and~}t\le k-2; \\
  \big\langle\boldsymbol{\alpha}^{s},\boldsymbol{\alpha}^{2k+2t-2-(q-1)}~(s\in\mathbf{\mathcal{S}}_{2k-2+t})\big\rangle,&\text{if~} h\ge 1\text{~and~}t+h\le k-1.\\
  \end{cases}\\
  =&\begin{cases}
  \big\langle\boldsymbol{\alpha}^{s},\boldsymbol{\alpha}+\eta\boldsymbol{\alpha}^{2k-1},\boldsymbol{1}+\eta^2\boldsymbol{\alpha}^{4k-3-q}~(\mathbf{\mathcal{S}}_{3k-3}\backslash\{0,1,2k-1\})\big\rangle,
  & \text{if~}h=0\text{~and~}t=k-1;\\
  \big\langle\boldsymbol{\alpha}^{s}~(s\in\mathbf{\mathcal{S}}_{2k-2+t})\big\rangle,&\text{if~}h=0\text{~and~}t\le k-2; \\
  &~~~\!\text{or~} h\ge 1\text{~and~}t+h\le k-1.\\
  \end{cases}\\
 	=&\begin{cases}
 	\big\langle\boldsymbol{\alpha}^{s},\boldsymbol{\alpha}+\eta\boldsymbol{\alpha}^{2k-1},\boldsymbol{1}+\eta^2\boldsymbol{\alpha}~(\mathbf{\mathcal{S}}_{3k-3}\backslash\{0,1,2k-1\})\big\rangle,
 	&\!\!\!\! \text{if~}q=4k-4,h=0\text{~and~}t=k-1;\\
 	\big\langle\boldsymbol{\alpha}^{s},\boldsymbol{\alpha}+\eta\boldsymbol{\alpha}^{2k-1}~(\mathbf{\mathcal{S}}_{3k-3}\backslash\{1,2k-1\})\big\rangle,
 	&\!\!\!\! \text{if~}q<4k-4,h=0\text{~and~}t=k-1;\\
 	\mathcal{RS}_{2k-1+t},&\!\!\!\!\text{if~}h=0\text{~and~}t\le k-2;\\
 	&~\!\text{or~} h\ge 1\text{~and~}t+h\le k-1.
 	\end{cases}
 	\end{align*} 
 }
 {\bf Case 3}. For $3\le k\le \frac{q-2t+1}{2}$, we have $2k+2t-2\le q-1$, thus $(3)$ holds. $\hfill\Box$\\
 
The following lemma is needed to calculate $\big(\mathcal{C}^2_{\mathcal{V}_{k,t,h,\eta}}\big)^{\perp}$. 
\begin{lemma}\label{q1} For any positive integer $l< q-1$, if $s_1\in\mathbf{\mathcal{S}}_l$ and $s_2\in\mathbf{\mathcal{S}}_l^{\perp}$, then 
	\begin{align}\label{fq1}
	\sum_{i=1}^{q}\alpha_i^{s_1}\alpha_i^{s_2}=0.
	\end{align}
\end{lemma}

{\bf Proof}. For any non-negative integer $l$, it is well-known that
\begin{align}\label{fq}
\sum_{i=1}^{q}\alpha^l=\begin{cases}
-1,\quad&\text{if~}q-1\mid l;\\
0,\quad&\text{otherwise}.
\end{cases}
\end{align}
Now by the definitions of $\mathbf{\mathcal{S}}_l$ and $\mathbf{\mathcal{S}}_l^{\perp}$, we have $$0\le s_1+s_2< 2(q-1) \text{~~and~~}s_1+s_2\neq q-1,$$ thus by $(\ref{fq})$, we obtain $(\ref{fq1})$. $\hfill\Box$\\

$\big(\mathcal{C}^2_{\mathcal{V}_{k,t,h,\eta}}\big)^{\perp}$ is determined in the following lemma.
\begin{lemma}\label{L3} For any  $3\le k\le \frac{q}{2}$, $t+h\le k-1$ and $\eta\in\mathbb{F}_q^{*}$, the following three assertions hold.\\
	
	$(1)$ If $\frac{q-t+1}{2}<k\le\frac{q}{2}$, then 
	\begin{align*}
	\big(\mathcal{C}^2_{\mathcal{V}_{k,t,h,\eta}}\big)^{\perp}=\{\boldsymbol{0}\}.
	\end{align*} 
	
	$(2)$ If $\frac{q-2t+1}{2}<k\le \frac{q-t+1}{2}$, then
	{\small	\begin{align*}
		&\big(\mathcal{C}^2_{\mathcal{V}_{k,t,h,\eta}}\big)^{\perp}\\
		=&\begin{cases}
		\big\langle\boldsymbol{\alpha}^{s} ,\boldsymbol{\alpha}^{q-2k}-\eta\boldsymbol{\alpha}^{q-2}+\eta^3\boldsymbol{\alpha}^{q-1}~(s\in\mathcal{S}_{3k-3}^{\perp})\big\rangle,
		&\!\!\!\! \text{if~}h=0,t=k-1\text{~and~}q=4k-4;\\
		\big\langle\boldsymbol{\alpha}^{s},\boldsymbol{\alpha}^{q-2k}-\eta\boldsymbol{\alpha}^{q-2}~ (s\in\mathcal{S}_{3k-3}^{\perp})\big\rangle,
		&\!\!\!\! \text{if~}h=0,t=k-1\text{~and~}q<4k-4;\\
		\mathcal{RS}_{q-2k-t+1},&\!\!\!\!\text{if~}h=0\text{~and~}t\le k-2,\\
		&\text{or~} h\ge 1\text{~and~}t+h\le k-1.
		\end{cases}
		\end{align*} 
	}

	$(3)$ If $3\le k\le \frac{q-2t+1}{2}$, then
	{\small
		\begin{align*}
		&\big(\mathcal{C}^2_{\mathcal{V}_{k,t,h,\eta}}\big)^{\perp}\\
		=&\begin{cases}
		\big\langle\boldsymbol{\alpha}^{s},\boldsymbol{\alpha}^{q-2k}+\boldsymbol{\alpha}^{q-4k+3}-\eta\boldsymbol{\alpha}^{q-2}-\eta^2\boldsymbol{\alpha}^{q-1}\big(s\in(\mathcal{S}_{3k-3}\cup\{4k-4\}
		)^{\perp}\big)\big\rangle,
		&\!\!\!\! \text{if~}h=0\text{~and~}t=k-1;\\
		\big\langle\boldsymbol{\alpha}^{s},\boldsymbol{\alpha}^{q-2k-2t+1}-\eta^2\boldsymbol{\alpha}^{q-1}\big(s\in(\mathcal{S}_{2k+t-2}\cup\{2k+2t-2\}
		)^{\perp}\big)\big\rangle,&\!\!\!\!\text{if~}h=0\text{~and~}t\le k-2; \\
		\big\langle\boldsymbol{\alpha}^{s}\big(s\in(\mathcal{S}_{2k+t-2}\cup\{2k+2t-2\}
		)^{\perp}\big)\big\rangle,&\!\!\!\!\text{if~} h\ge 1\text{~and~}t+h\le k-1.\\
		\end{cases}
		\end{align*} }
\end{lemma}	

{\bf Proof}. By Lemma \ref{L2}, we have the following three cases.\\

{\bf Case 1}. For $ \frac{q-t+1}{2}< k \le \frac{q}{2}$, we can obtain $(1)$  since $(\mathbb{F}_q^q)^{\perp}=\{\boldsymbol{0}\}$.\\

{\bf Case 2}. For $\frac{q-2t+1}{2}<k\le \frac{q-t+1}{2}$, we have the following three cases.\\

$(i)$ If $q=4k-4$ and $(h,t)=(0,k-1)$, then for any $s_1\in\mathcal{S}_{3k-3}\backslash\{0,1,2k-1\}$ and $s_2\in\mathcal{S}_{3k-3}^{\perp}$, by $(\ref{fq1})$ one has
\begin{align*}
&\sum_{i=1}^{q}\alpha_i^{s_1}\alpha_i^{s_2}=0,~~\sum_{i=1}^{q}(\alpha_i+\eta\alpha_i^{2k-1})\alpha_i^{s_2}=0,\\
&\sum_{i=1}^{q}(1+\eta^2\alpha_i)\alpha_i^{s_2}=0,~~
\sum_{i=1}^{q}\alpha_i^{s_1}(\alpha_i^{q-2k}-\eta\alpha_i^{q-2}+\eta^3\alpha_i^{q-1})=0,\\
&\sum_{i=1}^{q}(\alpha_i+\eta\alpha_i^{2k-1})(\alpha_i^{q-2k}-\eta\alpha_i^{q-2}+\eta^3\alpha_i^{q-1})=0,\\
&\sum_{i=1}^{q}(1+\eta^2\alpha_i)(\alpha_i^{q-2k}-\eta\alpha_i^{q-2}+\eta^3\alpha_i^{q-1})=0,
\end{align*}and then$$	\big\langle\boldsymbol{\alpha}^{s} ,\boldsymbol{\alpha}^{q-2k}-\eta\boldsymbol{\alpha}^{q-2}+\eta^3\boldsymbol{\alpha}^{q-1}~(s\in\mathcal{S}_{3k-3}^{\perp})\big\rangle\subseteq \big(\mathcal{C}^2_{\mathcal{V}_{k,t,h,\eta}}\big)^{\perp}.$$ Furthermore, by $(\ref{dd1})$, we have
$$\dim\Big(\big(\mathcal{C}^2_{\mathcal{V}_{k,t,h,\eta}}\big)^{\perp}\Big)=q-(3k-3)= \dim\Big(\big\langle\boldsymbol{\alpha}^{s} ,\boldsymbol{\alpha}^{q-2k}-\eta\boldsymbol{\alpha}^{q-2}+\eta^3\boldsymbol{\alpha}^{q-1}~(s\in\mathcal{S}_{3k-3}^{\perp})\big\rangle\Big),$$
Thus  \begin{align}\label{perp1}
 	\big(\mathcal{C}^2_{\mathcal{V}_{k,t,h,\eta}}\big)^{\perp}=	\big\langle\boldsymbol{\alpha}^{s},\boldsymbol{\alpha}^{q-2k}-\eta\boldsymbol{\alpha}^{q-2}+\eta^3\boldsymbol{\alpha}^{q-1}~(s\in\mathcal{S}_{3k-3}^{\perp})\big\rangle.
 \end{align}
 
$(ii)$ If $q=4k-4,h=0\text{~and~}t=k-1$, in the similar proof to that of $(\ref{perp1})$, we have
\begin{align*}
\big(\mathcal{C}^2_{\mathcal{V}_{k,t,h,\eta}}\big)^{\perp}=	\big\langle\boldsymbol{\alpha}^{s},\boldsymbol{\alpha}^{q-2k}-\eta\boldsymbol{\alpha}^{q-2}~(s\in\mathcal{S}_{3k-3}^{\perp})\big\rangle.
\end{align*}

$(iii)$ If $h=0\text{~and~}t\le k-2$, or $h\ge 1\text{~and~}t+h\le k-1$, by Lemma \ref{RS}, we have
\begin{align*}
	\big(\mathcal{C}^2_{\mathcal{V}_{k,t,h,\eta}}\big)^{\perp}=\mathcal{RS}_{q-2k-t+1}.
\end{align*}

So far, we can get $(2)$.\\

{\bf Case 3}. For $3\le k\le \frac{q-2t+1}{2}$, in the similar proof to that of {\bf Case 2}, we can obtain $(3)$.\\

	\begin{remark}\label{r32} Lemma \ref{L3} can be equivalently given from the perspective of linear spaces, that is 
		$$\big(\mathcal{C}^2_{\mathcal{V}_{k,t,h,\eta}}\big)^{\perp}=\mathcal{C}_{\mathcal{V}_{k,t,h,\eta}^{\perp}},$$
		where $\mathcal{V}_{k,t,h,\eta}^{\perp}$ is a  $\mathbb{F}_q$-linear space of $\mathbb{F}_q[x]$ and is defined as 

	$(1)$ for $\frac{q-t+1}{2}<k\le\frac{q}{2}$,
		\begin{align*}
		\mathcal{V}_{k,t,h,\eta}^{\perp}=\{0\};
		\end{align*}
	
	$(2)$ for  $\frac{q-2t+1}{2}<k\le \frac{q-t+1}{2}$, 	
	{\small	\begin{align*}
		\mathcal{V}_{k,t,h,\eta}^{\perp}
		=&\begin{cases}
		\bigg\{\sum\limits_{i=0}^{q-3k+1}a_i x^{i}\! +\! a(\eta^3 x^{q-1}\!-\!\eta x^{q-2}\!+\!x^{q-2k})\!~|\!~a,a_i\in\mathbb{F}_q\bigg\},
		&\!\!\!\! \text{if~}q=4k-4,h=0\text{~and~}t=k-1;\\
		\bigg\{\sum\limits_{i=0}^{q-3k+1}a_i x^{i}+ a(-\eta x^{q-2}+x^{q-2k})~|~a,a_i\in\mathbb{F}_q\bigg\},
		&\!\!\!\! \text{if~}q<4k-4,h=0\text{~and~}t=k-1;\\
		\big\{f(x)~|~\deg f(x)\le q-2k-t\big\},&\!\!\!\!\text{if~}h=0\text{~and~}t\le k-2,\\
		&\text{or~} h\ge 1\text{~and~}t+h\le k-1;
		\end{cases}
		\end{align*}}
	
	$(3)$ for $3\le k\le \frac{q-2t+1}{2}$,
	{\small
		\begin{align*}
		\mathcal{V}_{k,t,h,\eta}^{\perp}
		=&\begin{cases}
		\bigg\{  \sum\limits_{i=0}^{q-3k+1}a_i x^{i}+a_{q-4k+3}(x^{q-2k}-\!\eta x^{q-2}-\!\eta^2 x^{q-1})\! ~|~a_i\in\mathbb{F}_q\bigg\},
		& \text{if~}h=0\text{~and~}t=k-1;\\
		\bigg\{  \sum\limits_{i=0}^{q-2k-t}a_i x^{i}-a_{q-2k-2t+1}\eta^2 x^{q-1} ~|~a_i\in\mathbb{F}_q\bigg\},&\text{if~}h=0\text{~and~}t\le k-2; \\
		\bigg\{ \sum\limits_{\substack{i=0\\i\neq q-2k-2t+1}}^{q-2k-t}a_i x^{i} ~|~a_i\in\mathbb{F}_q\bigg\},&\text{if~} h\ge 1\text{~and~}t+h\le k-1.\\
		\end{cases}
		\end{align*} }
\end{remark}
	\section{Self-orthogonal GTRS Codes}
	In this section, we give a sufficient and necessary condition for a  GTRS code to be self-orthogonal. Furthermore, we construct some GTRS codes basing on this sufficient and necessary condition.
	\subsection{A sufficient and necessary condition for a GTRS code to be self-orthogonal}
	
	\begin{theorem}\label{t1}
	For $2k\le n\le q$ and $t+h\le k-1$, let $\boldsymbol{\alpha}=(\alpha_{i_1},\ldots,\alpha_{i_n})$ and $\boldsymbol{v}=(v_1,\ldots,v_n)\in(\mathbb{F}_q^n)^{*}$, then $\mathcal{C}_{\mathcal{V}_{k,t,h,\eta}}(\boldsymbol{\alpha},\boldsymbol{v})$ is self-orthogonal if and only if  there is a $f(x)\in\mathcal{V}_{k,t,h,\eta}^{\perp}$, such that $$f(\alpha_{i_j})=v_j^2 \quad (j=1,\ldots,n)\text{~~and~~}f(\beta)=0 \quad (\beta\in\mathbb{F}_q\backslash\{\alpha_{i_1},\ldots,\alpha_{i_n}\}).$$
	\end{theorem}

	{\bf Proof}. 
	By the definition, $\mathcal{C}_{\mathcal{V}_{k,t,h,\eta}}(\boldsymbol{\alpha},\boldsymbol{v})$ is self-orthogonal if and only if for any $$(v_1c_{1,i_1},\ldots,v_nc_{1,i_n}),~(v_1c_{2,i_1},\ldots,v_nc_{2,i_n})\in \mathcal{C}_{\mathcal{V}_{k,t,h,\eta}}(\boldsymbol{\alpha},\boldsymbol{v}),$$
	 we have
	\begin{align}\label{ddd}
	\sum_{j=1}^{n}(v_jc_{1,i_j})(v_jc_{2,i_j})=\sum_{j=1}^{n}v_j^2c_{1,i_j}c_{2,i_j}=0.
	\end{align}
	Note that $$v_j\neq 0~~(j=1,\ldots,n) \text{~~and~~}  
	(c_{1,i_1},\ldots,c_{1,i_n}),~(c_{2,i_1},\ldots,c_{2,i_n})\in \mathcal{C}_{\mathcal{V}_{k,t,h,\eta}}(\boldsymbol{\alpha},\boldsymbol{1}),$$ 
	$(\ref{ddd})$ is equavient to that
	there is a $\mathbf{c} = (c_1,\ldots,c_q )\in\mathbb{F}_q^q$ with $c_i = 0$ $(i\in\{1,\ldots,q\}\backslash\{i_1,\ldots,i_n\})$ and $v_{j}^{2}=c_{i_j}$ $(j=1,\ldots,n)$ such that {for~any~}$(c_{1,1},\ldots,c_{1,q}),~(c_{2,1},\ldots,c_{2,q})\in \mathcal{C}_{\mathcal{V}_{k,t,h,\eta}},$
	\begin{align*}
	\sum_{j=1}^{q}c_{j}(c_{1,i_j}c_{2,i_j})=0.
	\end{align*}
	Namely, there is a $\mathbf{c}=(c_1,\ldots,c_q)\in \big(\mathcal{C}^2_{\mathcal{V}_{k,t,h,\eta}}\big)^{\perp}$ with $Supp(\mathbf{c})=\{i_1,\ldots,i_n\}$ and $v_{j}^{2}=c_{i_j}$ $(j=1,\ldots,n)$.
	
	Now by the assumption $t+h\le k-1$ and Lemma $\ref{L3}$, we complete the proof. $\hfill\Box$
	\\
	

In Remark \ref{r32}, for $k>\frac{q+1-t}{2}$, since $\mathcal{V}_{k,t,h,\eta}^{\perp}=\{0\}$, thus we have

	\begin{corollary}\label{t3}
	For $t+h\le k-1$,  there is no self-orthogonal GTRS code with dimension  $$k>\frac{q-t+1}{2}.$$
	\end{corollary}

By Theorem \ref{t1}, it is easy to see that if $\mathcal{C}_{\mathcal{V}_{k,t,h,\eta}}(\boldsymbol{\alpha},\boldsymbol{v})$ is self-orthogonal, then there exists a $f(x)\in\mathcal{V}_{k,t,h,\eta}^{\perp}$ with $q-n$ pairwise distinct roots in $\mathbb{F}_q$. Thus we have
	\begin{corollary}\label{t4}
	For any $t+h\le k-1$ with $(h,t)\neq(0,k-1)$ and $3\le k\le \frac{q-t+1}{2}$, there is no self-orthogonal GTRS code with length $n< 2k+t$.
	\end{corollary}
\subsection{The existence for a self-orthogonal GTRS code}

In this subsetion, we assume $t+h\le k-1$, and then based on the sufficient and necessary condition for the GTRS code with $h+t\le k-1$ to be self-orthogonal, some self-orthogonal GTRS codes are constructed.
\subsubsection{ The self-orthogonal GTRS code with dimension $k$ $(\frac{q-2t+1}{2}<k<\frac{q-t+1}{2})$}
\begin{theorem}\label{tc1}
	For $\frac{q-2t+1}{2}<k<\frac{q-t+1}{2}$, $l\le\frac{q-2k-t}{2}$, $n=q-l$, and $\eta\in\mathbb{F}_q^{*}$. If $\boldsymbol{\alpha}=(\alpha_{i_1},\ldots,\alpha_{i_{n}})$ and $\boldsymbol{v}=(v_1,\ldots,v_n)$ with $v_j=\prod\limits_{\beta\in\mathbb{F}_{q}\backslash\{\alpha_{i_1},\ldots,\alpha_{i_{n}}\}}(\alpha_{i_j}-\beta)$, then  $\mathcal{C}_{\mathcal{V}_{k,t,h,\eta}}(\boldsymbol{\alpha},\boldsymbol{v})$ is self-orthogonal.
\end{theorem}

{\bf Proof}. Let $$f(x)=\Big(\prod\limits_{\beta\in\mathbb{F}_{q}\backslash\{\alpha_{i_1},\ldots,\alpha_{i_{n}}\}}(x-\beta)\Big)^2,$$ by the assumptions $\frac{q-2t+1}{2}<k<\frac{q-t+1}{2}$ and $l\le\frac{q-2k-t}{2}$, we have $$\deg f(x)\le q-2k-t,$$ and then $f(x)\in\mathcal{V}_{k,t,h,\eta}^{\perp}$. 

Furthermore, it is easy to see that
\begin{align*}
v_j^2=f(\alpha_{i_j})\quad (j=1,\ldots,n),
\end{align*}
and 
\begin{align*}
f(\beta)=0\quad (\beta\in\mathbb{F}_{q}\backslash\{\alpha_{i_1},\ldots,\alpha_{i_{q-l}}\}).
\end{align*}

So far, by Theorem \ref{t1}, $\mathcal{C}_{\mathcal{V}_{k,t,h,\eta}}(\boldsymbol{\alpha},\boldsymbol{v})$ is self-orthogonal. $\hfill\Box$\\

By Lemmas \ref{mnds}-\ref{nmds}, we have
\begin{corollary}
	By taking $(h,t)=(0,1)$ in Theorem \ref{tc1}, if $q$ is odd, then  $\mathcal{C}_{\mathcal{V}_{k,1,0,\eta}}(\boldsymbol{\alpha},\boldsymbol{v})$ is a self-orthogonal NMDS code.
\end{corollary}
\begin{theorem}\label{tc2}
	For any integer $m>2$, $q=2^m$, $\frac{q-2t+1}{2}<k<\frac{q-t+1}{2}$ and $ n\ge 2k+t$. If $\eta\in\mathbb{F}_q^{*}$, $\boldsymbol{\alpha}=(\alpha_{i_1},\ldots,\alpha_{i_n})$ and $\boldsymbol{v}=(v_1,\ldots,v_n)$ with $v_j=\prod\limits_{\beta\in\mathbb{F}_{2^m}\backslash\{\alpha_{i_1},\ldots,\alpha_{i_n}\}}(\alpha_{i_j}-\beta)^{2^{m-1}}$, then $\mathcal{C}_{\mathcal{V}_{k,t,h,\eta}}(\boldsymbol{\alpha},\boldsymbol{v})$ is self-orthogonal.
\end{theorem}

{\bf Proof}. Let $$f(x)=\prod\limits_{\beta\in\mathbb{F}_{2^m}\backslash\{\alpha_{i_1},\ldots,\alpha_{i_n}\}}(x-\beta),$$ by the assumptions $\frac{q-2t+1}{2}<k<\frac{q-t+1}{2}$ and $n\le2k+t$, we have $$\deg f(x)\le q-2k-t.$$  Thus $f(x)\in\mathcal{V}_{k,t,h,\eta}^{\perp}$. Furthermore, it is easy to see that
\begin{align*}
v_j^2=\prod_{\beta\in\mathbb{F}_{2^m}\backslash\{\alpha_{i_1},\ldots,\alpha_{i_n}\}}(\alpha_{i_j}-\beta)^{2^{m}}=f(\alpha_{i_j})\quad (j=1,\ldots,n),
\end{align*}
and 
\begin{align*}
f(\beta)=0\quad (\beta\in\mathbb{F}_{2^m}\backslash\{\alpha_{i_1},\ldots,\alpha_{i_n}\}).
\end{align*}

Now by Theorem \ref{t1}, $\mathcal{C}_{\mathcal{V}_{k,t,h,\eta}}(\boldsymbol{\alpha},\boldsymbol{v})$ is self-orthogonal. $\hfill\Box$\\

By Lemma \ref{mnds} and Remark \ref{r4}, we have
\begin{corollary}
By	taking $(h,t)=(0,1)$ in Theorem \ref{tc2}, if  $\eta\notin T_k(\boldsymbol{\alpha})$, then $\mathcal{C}_{\mathcal{V}_{k,1,0,\eta}}(\boldsymbol{\alpha},\boldsymbol{v})$ is a non-GRS self-orthogonal MDS code. Otherwise,  $\mathcal{C}_{\mathcal{V}_{k,1,0,\eta}}(\boldsymbol{\alpha},\boldsymbol{v})$ is a self-orthogonal NMDS code.
\end{corollary}
\subsubsection{ The Self-orthogonal GTRS code with dimension $k$ $(3\le k\le \frac{q-2t+1}{2})$}
\begin{theorem}\label{ct4}
	For prime $p$, positive integers $r$ and $m$ with $3\le r$ and $r\mid m$, if $q=p^m$, $n=p^r$,  $\eta\in\mathbb{F}_{p^m}^{*}$, $\mathbb{F}_{p^r}=\{\alpha_{i_1},\ldots,\alpha_{i_n}\}$ and  $\boldsymbol{\alpha}=(\alpha_{i_1},\ldots,\alpha_{i_n})$, then  $\mathcal{C}_{\mathcal{V}_{k,t,h,\eta}}(\boldsymbol{\alpha},\boldsymbol{1})$ is self-orthogonal if one of the following conditions holds.\\
	
	$(1)$  $t=1$ and $2k+t+2=p^{r}$;\\
	
	$(2)$  $t$ is odd with $t\ge 3$, and $2k+t=p^r$;\\
	
	$(3)$ $t=2$ and $2k+t+3=p^r$;\\
	
	$(4)$  $t$ is even with $t\ge 4$, and $2k+t+1=p^r$.
	
	
	
\end{theorem}

{\bf Proof}.  Let $$f(x)=\prod\limits_{\beta\in\mathbb{F}_{p^m}\backslash\mathbb{F}_{p^r}}(x-\beta),$$
then 
$$\deg f(x)=p^m-p^r\text{~~and~~} f(x)=\frac{x^{p^m}-x}{x^{p^r}-x}=\sum_{i=0}^{\frac{p^m-1}{p^r-1}-1}x^{i(p^r-1)}.$$

If one of the conditions $(1)$-$(4)$ holds, then we have $3\le k<\frac{p^m-t+1}{2}$ and

$$p^m-2k-2t+1\neq i(p^r-1)+1~\Big(i=0,\ldots,\frac{p^m-1}{p^r-1}-1\Big).$$ 
Thus the coefficient of $x^{p^m-2k-2t+1}$ in $f(x)$ is zero, and so $f(x)\in\mathcal{V}_{k,t,h,\eta}^{\perp}$.

Furthermore, we have
\begin{align*}
f(\alpha_{i_j})=\prod_{\beta\in\mathbb{F}_{p^m}\backslash\mathbb{F}_{p^r}}(\alpha_{i_j}-\beta)=\prod_{\beta\in\mathbb{F}_{p^m}\backslash\{\alpha_{i_j}\}}(\alpha_{i_j}-\beta)\Big(\prod_{\beta\in\mathbb{F}_{p^r}\backslash\{\alpha_{i_j}\}}(\alpha_{i_j}-\beta)\Big)^{-1}=1,
\end{align*}
and 
\begin{align*}
f(\beta)=0\quad (\beta\in\mathbb{F}_{p^m}\backslash\mathbb{F}_{p^r}).
\end{align*}

Now by Theorem $\ref{t1}$, we have the desired results. $\hfill\Box$\\

\begin{theorem}\label{ct5}For any positive integers $r$, $m$ with $3\le r$ and $r\mid m$, if $n=2^r-1$, $q=2^m$, $\eta\in\mathbb{F}_{2^m}^{*}$, $\mathbb{F}_{p^r}^{*}=\{\alpha_{i_1},\ldots,\alpha_{i_n}\}$, $\boldsymbol{\alpha}=(\alpha_{i_1},\ldots,\alpha_{i_n})$ and $v_j=\prod\limits_{\beta\in\mathbb{F}_{2^m}\backslash\mathbb{F}_{2^r}^{*}}(\alpha_{i_j}-\beta)^{2^{m-1}}\quad (j=1,\ldots,n)$, 
then  $\mathcal{C}_{\mathcal{V}_{k,t,h,\eta}}(\boldsymbol{\alpha},\boldsymbol{v})$ is self-orthogonal if one of the following conditions holds.\\

	$(1)$ $t=2$ and $2k+t+3=2^{r}-1$;\\
	
	$(2)$ $t$ is even with $t\ge 4$, and $2k+t+1=2^{r}-1$;\\  
	
	$(3)$  $t=1$ and $2k+t+2=2^r-1$;\\
	
	$(4)$  $t$ is odd, with $t\ge 3$ and $2k+t=2^r-1$.	
\end{theorem}

{\bf Proof}.  Let $$f(x)=\prod\limits_{\beta\in\mathbb{F}_{2^m}\backslash\mathbb{F}_{2^r}^{*}}(x-\beta),$$
then 
$$\deg f(x)=2^m-2^r+1\text{~~and~~} f(x)=\frac{x^{2^m}-x}{x^{2^r-1}-1}=\sum_{i=0}^{\frac{2^m-1}{2^r-1}-1}x^{i(2^r-1)+1}.$$

If one of the conditions $(1)$-$(4)$ holds, then we have $3\le k<\frac{2^m-t+1}{2}$ and 

$$2^m-2k-2t+1\neq i(2^r-1)+1~~\Big(i=0,\ldots,\frac{2^m-1}{2^r-1}-1\Big).$$ 
Thus the coefficient of $x^{2^m-2k-2t+1}$ in $f(x)$ is zero, and so $f(x)\in\mathcal{V}_{k,t,h,\eta}^{\perp}$.

Furthermore, one has
\begin{align*}
v_j^2=\prod_{\beta\in\mathbb{F}_{2^m}\backslash\mathbb{F}_{2^r}^{*}}(\alpha_{i_j}-\beta)^{2^{m}}=f(\alpha_{i_j})\quad (j=1,\ldots,n),
\end{align*}
and 
\begin{align*}
f(\beta)=0\quad (\beta\in\mathbb{F}_{2^m}\backslash\mathbb{F}_{2^r}^{*}).
\end{align*}

Now, by Theorem $\ref{t1}$, we have the desired results. $\hfill\Box$\\

By Lemma \ref{lmds} and Remark \ref{r4}, we have
\begin{corollary}
	By taking $\eta\in\mathbb{F}_p^{m}\backslash\mathbb{F}_{p^r}$ in Theorems \ref{ct4}-\ref{ct5},   $\mathcal{C}_{\mathcal{V}_{k,t,h,\eta}}(\boldsymbol{\alpha},\boldsymbol{v})$ is a non-GRS self-orthogonal MDS code.
\end{corollary}
\section{Conclusions and Further Study}
In this paper, we give a sufficient and necessary condition for the GTRS code with $h+t\le k-1$ to be self-orthogonal. And then, based on this sufficient and necessary condition,  some of  non-GRS self-orthogonal MDS codes or NMDS codes are constructed from these GTRS codes.

In Corollary \ref{t4}, we show that for any $t+h\le k-1$ with $(h,t)\neq(0,k-1)$ and $3\le k\le \frac{q-t+1}{2}$, there is no self-orthogonal GTRS code with length $n< 2k+t$. However,  under the condition $(h,t)=(0,k-1)$, we can not obtain the similar result, or construct a self-orthogonal GTRS code with length $n<2k+t$.

Furthermore, for the case $h+t\ge k$, it is interesting to give a sufficient and necessary condition for the GTRS code to be self-orthogonal.

\end{document}